%% file: RedWine.tex
\documentclass[seceq]{ptptex}
\usepackage{graphicx}
\usepackage{wrapft}
\usepackage{epsf}
\usepackage{color}
\usepackage{amsmath,amssymb,color,graphics,amscd,amsfonts}

\newcommand{\Tr}{\mathrm{Tr}}

\title{
String Tension and String Susceptibility \\
in two-dimensional generalized Weingarten model
}

\author{
Masanori \textsc{Hanada}\footnote{
E-mail:~hanada@gauge.scphys.kyoto-u.ac.jp
}
and
Fukuichiro \textsc{Kubo}\footnote{
E-mail:~kubo@gauge.scphys.kyoto-u.ac.jp} 
}

\inst{
Department of Physics, Kyoto University,
Kyoto 606-8502, Japan 
}

\recdate{December 27, 2006}

\abst{
We study the two-dimensional generalized Weingarten model 
reduced to a point, which interpolates between the
reduced Weingarten model and the large-$N$ gauge theory. 
We calculate the expectation value of the Wilson loop 
using a Monte Carlo method and determine the string tension 
and string susceptibility. 
The numerical result suggests that 
the string susceptibility approaches $-2$ 
in a certain parametric region, 
which implies that the branched-polymer configurations  
are suppressed. 
}

\begin{document}

\maketitle

\section{Introduction}

In this paper, we study a model defined by the action
\begin{eqnarray}
  S=
  -\beta N\sum_{\mu\neq\nu}^d \Tr
  \left(A_\mu^\dagger A_\nu^\dagger A_\mu A_\nu\right)
  +
  \alpha N\sum_{\mu=1}^d \Tr
  \left(
    A_\mu^\dagger A_\mu-1+\frac{1}{\alpha}
  \right)^2,  
  \label{action}
\end{eqnarray}
where $A_\mu$~$(\mu=1,2,\cdots,d)$ are complex $N\times N$ matrices. 
(In the actual numerical calculation describe in \textsection~ 
\ref{subsec:numerical results}, we consider only the case $d=2$.)
This model\footnote{
This model is the zero-dimensional reduced version of 
the ``interpolating model'' proposed in Ref.~\citen{IMS}.  
For $\alpha>1$, this model is equivalent to that
studied in Ref.~\citen{HKKK},
\begin{eqnarray}
  S=
  -\gamma N\sum_{\mu\neq\nu}^d \Tr
  \left(X_\mu^\dagger X_\nu^\dagger X_\mu X_\nu\right)
  +
  \kappa N\sum_{\mu=1}^d \Tr
  \left(
    X_\mu^\dagger X_\mu-1
  \right)^2,  
  \label{action_HKK}
\end{eqnarray}
with the relation  
\begin{eqnarray}
  A_\mu=\sqrt{1-\frac{1}{\alpha}}X_\mu, 
  \qquad
  \kappa=\alpha\left(1-\frac{1}{\alpha}\right)^2, 
  \qquad
  \beta=\frac{\gamma}{\left(1-\frac{1}{\alpha}\right)^2}. 
\end{eqnarray}
} interpolates \cite{IMS} between
the reduced Weingarten model \cite{EK3} 
and the large-$N$ reduced $U(N)$ gauge theory.\cite{EK1}  
We regard the model (\ref{action}) as describing an ensemble of 
random surfaces and study whether it can describe 
the Nambu-Goto string.

The original Weingarten model \cite{Weingarten} was proposed
as a nonperturbative description of the Nambu-Goto string.
This model is defined as follows. 
Consider a $d$-dimensional square lattice ${\mathbb Z}^d$ and
introduce a complex $N\times N$ matrix $A_{x,\mu}$
for each link connecting the sites $x$ and $x+\hat{\mu}$
in such a way that $A_{x+\hat{\mu},-\mu}=A_{x,\mu}^\dagger$. 
Then the action of the Weingarten model is given by 
\begin{eqnarray}
  S_W
  =
  -N\beta\sum_x\sum_{\mu\neq\nu}
  \Tr\left(
    A_{x,\mu}A_{x+\hat{\mu},\nu}
    A^\dagger_{x+\hat{\nu},\mu}A^\dagger_{x,\nu}
  \right)
  +N\sum_x\sum_{\mu=1}^d 
  \Tr\left(
    A_{x,\mu}^\dagger A_{x,\mu}
  \right).
  \label{action:original Weingarten}
\end{eqnarray}
The partition function is given by 
\begin{eqnarray}
  Z_W=\int dm_N \exp\left(-S_W\right), 
\end{eqnarray}
where the measure $dm_N$ is defined by 
\begin{eqnarray}
  dm_N
  =
  \prod_{x,\mu}\prod_{i,j}
  \left(
    \frac{N}{\pi}
    d\left[{\rm Re} (A_{x,\mu})_{ij}\right]
    d\left[{\rm Im} (A_{x,\mu})_{ij}\right]
  \right). 
\end{eqnarray}
Let $C_i$ represent closed contours on the lattice. 
Then, multiplying $A_\mu$ along $C_i$ and taking the trace, 
we obtain Wilson loops $w(C_i)$. 
The correlator of $w(C_i)$, defined by 
\begin{eqnarray}
  W(C_1,\cdots,C_n)
  =
  \frac{1}{Z_W}\int dm_N \exp\left(-S_W\right)
  \frac{1}{N}w(C_1)\cdots\frac{1}{N}w(C_n),
  \label{Wilson loop correlator0}
\end{eqnarray}
is evaluated as
\begin{eqnarray}
  W(C_1,\cdots,C_n)
  =
  N^{2-2n}\sum_{s\in S(\{C_i\})}
  \exp\left(
    -a(s)\log\beta^{-1}-h(s)\log N^2
  \right), 
  \label{Wilson loop correlator}
\end{eqnarray}
where $S(\{C_i\})$ is the set of surfaces on the lattice 
whose boundary is $C_1\cup\cdots\cup C_n$, 
$a(s)$ is the area of the surface $s$, and 
$h(s)$ is the number of handles of $s$. 
If we regard $\log\beta^{-1}$ and $\frac{1}{N^2}$ 
as the string tension and the string coupling, respectively,  
then Eq.~(\ref{Wilson loop correlator}) can be interpreted as 
the sum of random surfaces weighted by the Nambu-Goto action. 

Next, let us consider the reduced Weingarten model \cite{EK3}, whose
action is given by
\begin{eqnarray}
  S_{RW}
  =
  -N\beta\sum_{\mu\neq\nu}
  \Tr\left(
    A_{\mu}A_{\nu}
    A^\dagger_{\mu}A^\dagger_{\nu}
  \right)
  +N\sum_{\mu=1}^d 
  \Tr\left(
    A_{\mu}^\dagger A_{\mu}
  \right). 
  \label{action:reduced Weingarten}
\end{eqnarray}
This action is invariant under the $U(1)^d$ transformation 
\begin{eqnarray}
  A_\mu\to e^{i\theta_\mu}A_\mu. 
\end{eqnarray}
If this symmetry is not broken spontaneously in the large-$N$ limit,
the correlators of the Wilson loops
in this model are identical to those of the
original Weingarten model, (\ref{action:original Weingarten}). 
Because the reduced Weingarten model has only $d$ matrices, 
numerical calculations are more tractable.
This model was studied numerically \cite{KO} in the cases $d=2,3$, 
and it was shown that the Weingarten model does not describe 
smooth surfaces but, rather, branched polymers \cite{Kawai}. 

One possibility to overcome this difficulty 
is to consider the modified action (\ref{action}). 
This action is motivated by the following observations. 
First, this model interpolates \cite{IMS} between
the original Weingarten model 
($\alpha=0$)\footnote{
We need the redefinitions
\begin{eqnarray}
  A_\mu^{\mbox{(original)}}
  =
  \sqrt{2}A_\mu^{\mbox{(modified)}}, 
  \qquad
  \beta^{\mbox{(original)}}
  =
  \frac{1}{4}\beta^{\mbox{(modified)}}. 
\end{eqnarray}
} and the reduced $U(N)$ gauge theory \cite{EK1}($\alpha=\infty$).
Because both of those models are thought to be related 
to string theory, it is natural to consider the intermediate region. 
In this region, this model allows a lattice string interpretation
similar to that of the original Weingarten model, 
because the relation (\ref{Wilson loop correlator}) holds also in this
model, as long as the surface $s$ does not intersect itself.
Second, the action (\ref{action}) becomes a set of $d$ copies of 
a complex one-matrix model with a double-well potential. 
In the case of the Hermitian matrix model, 
we can describe a type 0B string 
by flipping the sign of the double-well potential \cite{TT}.
Therefore we conjecture that also in the case of the Weingarten model, 
a worldsheet supersymmetry can be introduced by modifying the potential,
and this may prevent the worldsheet from becoming branched polymer. 

This model has been solved analytically only in the special cases
$\beta=0$ \cite{IMS} and $\alpha=\infty$ for $d=2$.\cite{PR,GW} 
In Ref.~\citen{HKKK}, the parametric region $\alpha\gtrsim1$ 
is studied. 
For sufficiently large $\alpha$, there are 
$d$ phase transitions that correspond to 
the partial breakdowns of $U(1)^d$ symmetry.  
For $d\ge 3$, these phase transitions smoothly approach
the known phase transitions of the large-$N$ reduced 
$U(N)$ gauge theory \cite{NN,HNT} as $\alpha\to\infty$.  
For sufficiently small $\alpha$ 
($\alpha\lesssim 2$ in the case $d=2$), 
these two transitions seem to occur simultaneously. 
In this paper, we study the parametric region $\alpha\lesssim 1$  
in detail in the case $d=2$. 
For a technical reason, we study the maximally twisted model.   

The organization of this paper is as follows.
In \textsection~\ref{sec:theoretical background},  
we present theoretical preliminaries. 
In \textsection~\ref{subsec:numerical results},  
we present the numerical result. 
We study the phase diagram in \textsection~\ref{sec:Phase diagram} 
and then determine the string tension and string susceptibility 
in \textsection~\ref{sec:Wilson loop}. 
Section~\ref{sec:Discussion} is devoted to conclusions 
and discussion of future directions. 
In Appendix \ref{appendix:untwieted} we comment on numerical results 
concerning the generalized Weingarten model without a twist.
\section{Theoretical preliminaries
}\label{sec:theoretical background}
\subsection{Parametric region of the generalized Weingarten model
}\label{subsec:parametric region}

We begin by considering the parametric region in which
the model defined by the action (\ref{action}) is well-defined. 
As shown in REf.~\cite{HKKK}, the action is bounded from below 
if and only if 
\begin{eqnarray}
  \beta \alpha^{-1} \le \frac{1}{d-1}.  
  \label{bound}
\end{eqnarray}
This can be seen as follows. 
In order to determine whether the action is bounded, 
it is sufficient to consider the quartic term,
\begin{eqnarray}
  S|_{\mathrm{quartic}}=
  -\beta N\sum_{\mu\neq\nu}^d \Tr
  \left(A_\mu^\dagger A_\nu^\dagger A_\mu A_\nu\right)
  +
  \alpha N\sum_{\mu=1}^d \Tr
  \left(A_\mu^\dagger A_\mu A_\mu^\dagger A_\mu\right).  
  \label{quartic}
\end{eqnarray}
Then, using the inequality 
\begin{eqnarray}
  2\mathrm{Re}\  \mathrm{Tr}\left( A B^\dagger\right) \le \Tr \left(A A^\dagger\right) 
  + \mathrm{Tr} \left(B B^\dagger\right),  
\end{eqnarray}
we have 
\begin{eqnarray}
  2\mathrm{Re}\  \mathrm{Tr}\left( A_\mu A_\nu A_\mu^\dagger A_\nu^\dagger
  \right) 
  &\le& \mathrm{Tr} \left(A_\mu^\dagger A_\mu A_\nu A_\nu^\dagger\right)
          +\left(A_\mu A_\mu^\dagger A_\nu^\dagger A_\nu\right) 
          \nonumber\\
  &\le& \mathrm{Tr} \left(A_\mu^\dagger A_\mu A_\mu^\dagger A_\mu\right)
          +\left(A_\nu^\dagger A_\nu A_\nu^\dagger A_\nu\right).
\end{eqnarray}
Summing over the spacetime subscripts, we obtain
\begin{eqnarray}
  \sum_{\mu\neq\nu}^d \mathrm{Tr}
  \left(A_\mu^\dagger A_\nu^\dagger A_\mu A_\nu\right)
  \le
  (d-1)\sum_{\mu=1}^d \mathrm{Tr}
  \left(A_\mu^\dagger A_\mu \right)^2.
\end{eqnarray} 
Combining this relation with (\ref{quartic}), we obtain the bound   
(\ref{bound}). 
This bound is indeed realized in the case of the unit matrix, for example.

For $\beta \alpha^{-1}>\frac{1}{d-1}$, 
although the action is not bounded from below, 
the model still can be well-defined for large-$N$.\cite{EK3} 
In the original Weingarten model, the free energy per unit volume, 
$F(N,\beta)$, is given by 
\begin{eqnarray}
  \frac{1}{N^2}F(N,\beta)
  =
  \sum_{g=0}^\infty\sum_{A=0}^\infty
  \frac{n_g(A)}{N^{2g}}\beta^A, 
\end{eqnarray}
where $n_g(A)$ represents the number of closed surfaces with genus $g$ 
and area $A$. By taking the planar limit $N\to\infty$, with $\beta$ 
kept fixed,  we obtain 
\begin{eqnarray}
  \frac{1}{N^2}F(N,\beta)
  =
  \sum_{A=0}^\infty
  n_0(A)\beta^A. 
\end{eqnarray}
In Refs.~\citen{EK81} and \citen{DFJ83}, it is
argued that $n_0(A)$ behaves as 
\begin{eqnarray}
  n_0(A)\sim A^{b-1} k^A 
 \label{number_of_random_surface_without_boundary}
\end{eqnarray}
for sufficiently large $A$, where $b$ and $k$ are universal 
and regularization-dependent constants, respectively.  
Then, we have 
\begin{eqnarray}
  \frac{1}{N^2}F(N,\beta)
  \sim
  \sum_{A=0}^\infty
  A^{b-1}(k\beta)^A,  
\end{eqnarray}
and this quantity is finite for $\beta<\beta_c=\frac{1}{k}$. 
In the same way, it can be shown that
the expectation value of the Wilson loop 
is also finite. Therefore, the original Weingarten model 
is well-defined for $\beta<\beta_c$ in the large-$N$ limit.\cite{EK3} 

In numerical simulations, obviously we cannot make $N$ infinite.  
However, for large enough $N$, there is a metastable state 
corresponding to the planar limit. The ``lifetime'' of this metastable 
state becomes longer as $N$ increases.\cite{KO}    

For generic $\alpha$, the measure in the strong coupling expansion 
changes from the Gaussian one in the original Weingarten model as 
\begin{eqnarray}
  \exp\left(-N\sum_{\mu=1}^d \Tr A_{\mu}^\dagger A_{\mu}\right)
  dm_N
  \ 
  \longrightarrow
  \ 
  \exp\left(
    -
    \alpha N\sum_{\mu=1}^d \Tr
    \left(A_\mu^\dagger A_\mu-1+\frac{1}{\alpha}\right)^2
  \right)
  dm_N. 
\end{eqnarray}
Because we cannot evaluate the strong coupling expansion exactly, 
we simply {\it assume} that the effect of the change of the measure 
can be absorbed into changes of  $b$ and $\beta_c$:  
\begin{eqnarray}
  b,\ \beta_c
  \ 
  \longrightarrow 
  \ 
  b(\alpha),\ \beta_c(\alpha). 
  \label{ansatz:suppression of number of random surface}
\end{eqnarray}
If this assumption is correct, then the planar limit exists 
for $\beta<\beta_c(\alpha)$ also in this case. 
As we see in \textsection~\ref{subsec:numerical results}, 
the numerical data seem to be consistent with this assumption. 
\subsection{String tension and string susceptibility
}\label{subsec:susceptibility}
First, let us consider the case of the original Weingarten model. 
Then, the number of planar random 
surfaces on a lattice with boundary $C$ and area $A$  is given by 
\cite{EK81,DFJ83} 
\begin{eqnarray}
  n_0(A;C)
  \sim A^b k^A 
  =
  A^b \beta_c^{-A}
  \label{number_of_random_surface_with_boundary}
\end{eqnarray}
for large $A$. 
Note that $n_0(A;C)$ is $A$ times larger than $n_0(A)$,
because there is a degree of freedom corresponding to 
the choice of the location of a puncture. 
Then, the expectation value of the Wilson loop is evaluated as 
\begin{eqnarray}
  W(C)
  &=&
  \sum n_0(A;C)\beta^{A}
  \nonumber\\
  &\sim&
  \mbox{const.}\times
  \sum_{A\ge A_0} A^b (\beta/\beta_c)^A
  \nonumber\\
  &\sim&
  W_c(C)-
  \mbox{const.}\times
  \left|\beta-\beta_c\right|^{-b-1}
  +
  \cdots,  
  \label{eq:summation_of_random_surface}
\end{eqnarray}
where $W_c(C)=\left.W(C)\right|_{\beta=\beta_c}$ and 
$A_0$ represents the minimum area surrounded by $C$. 
Because the relation 
(\ref{number_of_random_surface_with_boundary}) can hold 
only for sufficiently large $A$, 
only the leading order singularity in the limit $\beta\to\beta_c$ 
is reliable in the expression (\ref{eq:summation_of_random_surface}).  

We can readily determine $b$ from (\ref{eq:summation_of_random_surface}).
In Refs.~\citen{DFJ84} and \citen{KO}, 
$b$ is determined to be $-1.5$. 
However, as we see later in this section, 
branched polymers dominate 
the path integral if $b>-2$ \cite{Kawai}.
Furthermore, the string tension is finite even at $\beta=\beta_c$ 
\cite{DFJ84}, 
although, in order to take the continuum limit, the string tension 
must approach zero.   
For these reasons, the original Weingarten model does not allow 
a meaningful continuum limit. 

Next let us consider the parametric region $\alpha>0$. 
In this case, if we assume the number of random surfaces 
is changed effectively  
as Eq.~(\ref{ansatz:suppression of number of random surface}), 
then we can expect the following behavior at $\beta\sim\beta_c(\alpha)$:  
\begin{eqnarray}
  W(C)
  &\sim&
  \sum_{A\ge A_0}\ A^{b(\alpha)} (\beta/\beta_c(\alpha))^A. 
  \nonumber\\
  &\sim&
  W_c(C)-
  \mbox{const.}\times
  \left|\beta-\beta_c(\alpha)\right|^{-b(\alpha)-1}
  +
  \cdots. 
  \label{eq:expectation_value_of_Wilson_loop_general_kappa}
\end{eqnarray}
In \textsection~\ref{subsec:numerical results} 
we determine the value of the ``string susceptibility'',
$b(\alpha)$, on the basis of this relation.  

\begin{wrapfigure}{1}{6.6cm}
    \qquad\qquad \scalebox{0.5}{
    \includegraphics{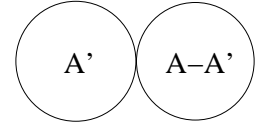}}
    \caption{A surface with a pinch. 
    }\label{fig:kubire}
\end{wrapfigure}
In the latter part of this section, following Ref.~\citen{Kawai} 
we show that smooth surfaces 
(resp., branched polymers) dominate 
the path integral if the string susceptibility $b$ is 
smaller (resp., larger) than $-2$. 
The number of planar surfaces with 
sufficiently large area $A$ is given by 
Eq.~(\ref{number_of_random_surface_without_boundary}).  
Now, let us assume that smooth surfaces are dominant for a given $b$. 
Then, the number of surfaces with a pinch (Fig.~\ref{fig:kubire}) 
is 
\begin{eqnarray}
  \sum_{A^\prime=0}^A n_0(A^\prime;C)n_0(A-A^\prime;C)
  \sim
  k^A  \sum_{A^\prime=0}^A
  A^{\prime b}(A-A^\prime)^{b}
  \sim
  k^A A^{2b+1},  
\end{eqnarray}
where $C$ represents the punctures to be connected. 
In order for the smooth surfaces to dominate the pathintegral, 
the number of smooth surfaces $n_0(A)$ 
must be larger than the number of surfaces with a pinch. 
Therefore, we have the following relation:  
\begin{eqnarray}
  n_0(A)\sim k^A A^{b-1}> k^A A^{2b+1}
  \Leftrightarrow 
  b<-2.  
\end{eqnarray}
By contrast, if $b>-2$, surfaces with more pinches contribute more, 
and hence branched polymers dominate the path integral. 
\section{Numerical results for the two-dimensional 
generalized Weingarten model with maximal twist}\label{subsec:numerical results}
In this section, we present the numerical result for 
the two-dimensional generalized Weingarten model. 
In order to determine the string susceptibility 
using the ansatz 
(\ref{eq:expectation_value_of_Wilson_loop_general_kappa}), 
we need to study the metastable region. 
Therefore, in order to suppress the tunneling effect, \cite{KO} 
we studied the maximally twisted reduced model \cite{GO82}. 
This is obtained by making the replacements  
\begin{eqnarray}
  \beta\to -\beta, 
  \qquad
  W[m,n]\to (-)^{mn}W[m,n],  
\end{eqnarray}
where $W[m,n]$ represents the expectation value of 
an $m\times n$ rectangular Wilson loop. 
\subsection{Phase diagram}\label{sec:Phase diagram}
\begin{figure}[b]
  \begin{center}
    \scalebox{0.5}{\input{phase_small_alpha.pstex_t}}
    \caption{
      Phase diagram of the two-dimensional generalized Weingarten
      model {\it with a twist} 
      for $\alpha\lesssim 1$. 
    }\label{phase diagram: small alpha with twist}
  \end{center}
\end{figure}
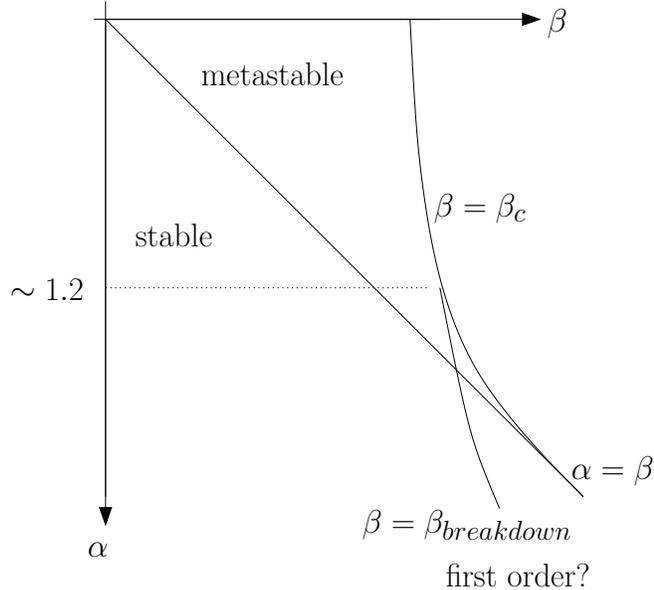

A schematic picture of the phase diagram for 
the maximally twisted reduced model is 
displayed in Fig.~\ref{phase diagram: small alpha with twist}. 
The line $\beta=\beta_c$ represents the boundary of 
the metastable region; i.e.,   
for $\beta<\beta_c$, 
the ``lifetime'' of the metastable state becomes longer 
as $N$ becomes larger. 
As the value of $\alpha$ increases, this line 
approaches the boundary of the stable region, $\alpha=\beta$. 
For $\alpha\sim 1.2$, a phase transition takes place at 
$\beta=\beta_{\mathrm{breakdown}}$.  
The values of $\beta_{\mathrm{breakdown}}$ 
seem to diverge as $\alpha\to\infty$. 
This is consistent with the analytic result at $\alpha=\infty$ 
(i.e., the unitary limit), according to
which no first-order or second-order phase 
transition exists.     

A remark is in order here. 
As shown in Appendix \ref{appendix:untwieted}, 
$\beta_{breakdown}$ seems to coincide with $\beta_1$ in 
untwisted model. This represents the breakdown of 
the $U(1)^2$ symmetry. Furthermore, the expectation value 
of the Wilson loop seems to coincide
with that in the untwisted model
not only for $\beta<\beta_\mathrm{breakdown}=\beta_1$ but also for
$\beta>\beta_\mathrm{breakdown}=\beta_1$.  
Therefore, it is plausible that the phase transition at 
$\beta=\beta_\mathrm{breakdown}$ corresponds to the breakdown of 
the $U(1)^2$ symmetry. If this is indeed the case, 
then the large-$N$ reduction\cite{EK3,EK1} cannot be applied 
for $\beta>\beta_{breakdown}$ and hence
the lattice-string interpretation
would not be possible for this model
in this parametric region. 
\subsection{Wilson loop}\label{sec:Wilson loop}
Let $W[m,n]$ be the expectation value of the $m\times n$ 
rectangular Wilson loop. 
We calculated the values of $W[m,n]$ with $m,n=1,\cdots,5$. 
For each $\alpha$ and $\beta$,  
we fitted the data numerically by using the ansatz 
\begin{eqnarray}
  W[m,n]= 
  c\cdot p^{2(m+n)}\cdot e^{-T\cdot mn},  
 \label{eq:arealaw}
\end{eqnarray}
where $c,p$ and $T$ are constants. 
There is a subtlety here: The numerical result suggests that 
the effect of the perimeter $p$ depends not only on $\alpha$ 
but also on $\beta$. Therefore, we expect the behavior described by 
Eq.~(\ref{eq:expectation_value_of_Wilson_loop_general_kappa}) 
only when the loop is large enough. 
In numerical simulations, because we can only study small loops, 
we should take the perimeter effect into account. 
Then, we expect 
\begin{eqnarray}
  p^{-L}W[m,n]
  \equiv
  W^\prime[m,n]
  \sim
  W^\prime_c[m,n]-
  \mbox{const.}\times
  \left|\beta-\beta_c(\alpha)\right|^{-b(\alpha)-1}
  +
  \cdots 
  \label{eq:expectation_value_of_Wilson_loop_with_perimeter_correction}
\end{eqnarray}
instead of Eq.~(\ref{eq:expectation_value_of_Wilson_loop_general_kappa}).  
As we see below, this ansatz appears to be consistent
with the numerical data. 

For $N=100$, only the loops satisfying $mn<10$ 
seem to converge 
(see Figs.~\ref{Fig:N dependence of W[1,1]} 
and \ref{Fig:N dependence of W[5,5]}). 
For this reason, we use only such values  
to extract the string tension and the string susceptibility.

\begin{figure}[t]
 \parbox{\halftext}{
      \scalebox{0.7}{\includegraphics{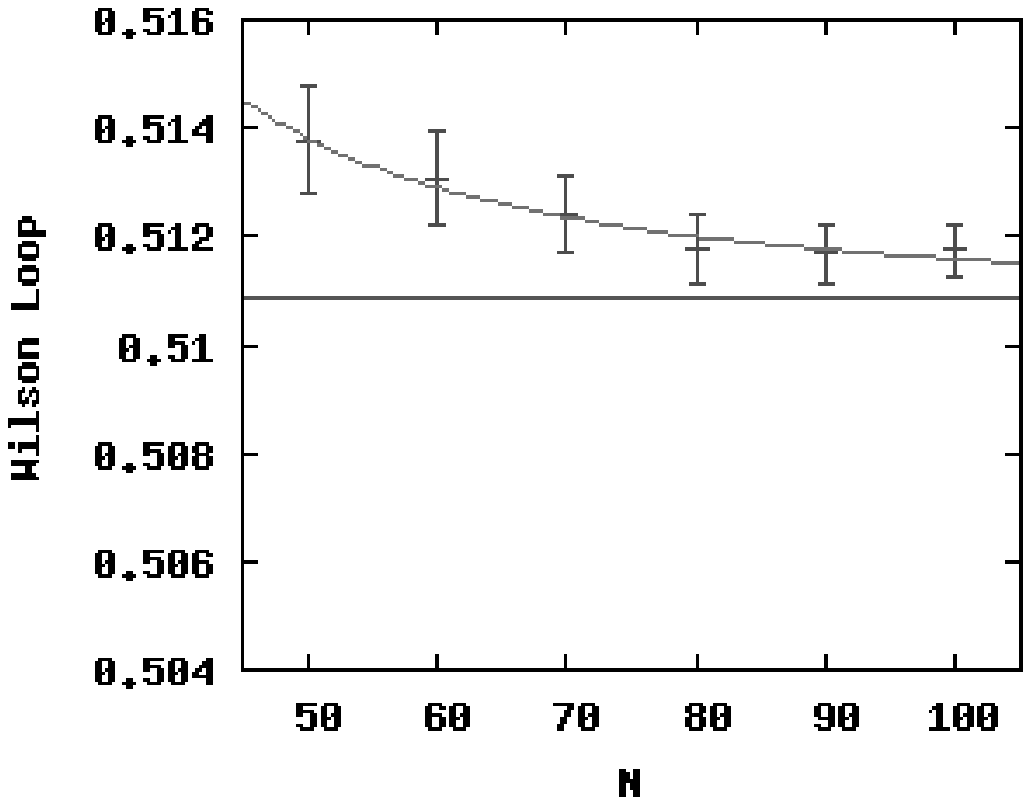}}
      \caption{
        $W[1,1]$ for $\alpha=3.73,\ \beta=0.4$. 
        The horizontal line represents the estimated value 
        in the large-$N$ limit, obtained by fitting the form 
        $W[1,1]=c_1+c_2 N^{c_3}$, where $c_1,c_2$ and $c_3$ 
        are constants. 
        We find good convergence at $N=100$. 
      }\label{Fig:N dependence of W[1,1]}
 }
 \hspace{8mm}
 \parbox{\halftext}{ 
        \scalebox{0.35}{\includegraphics{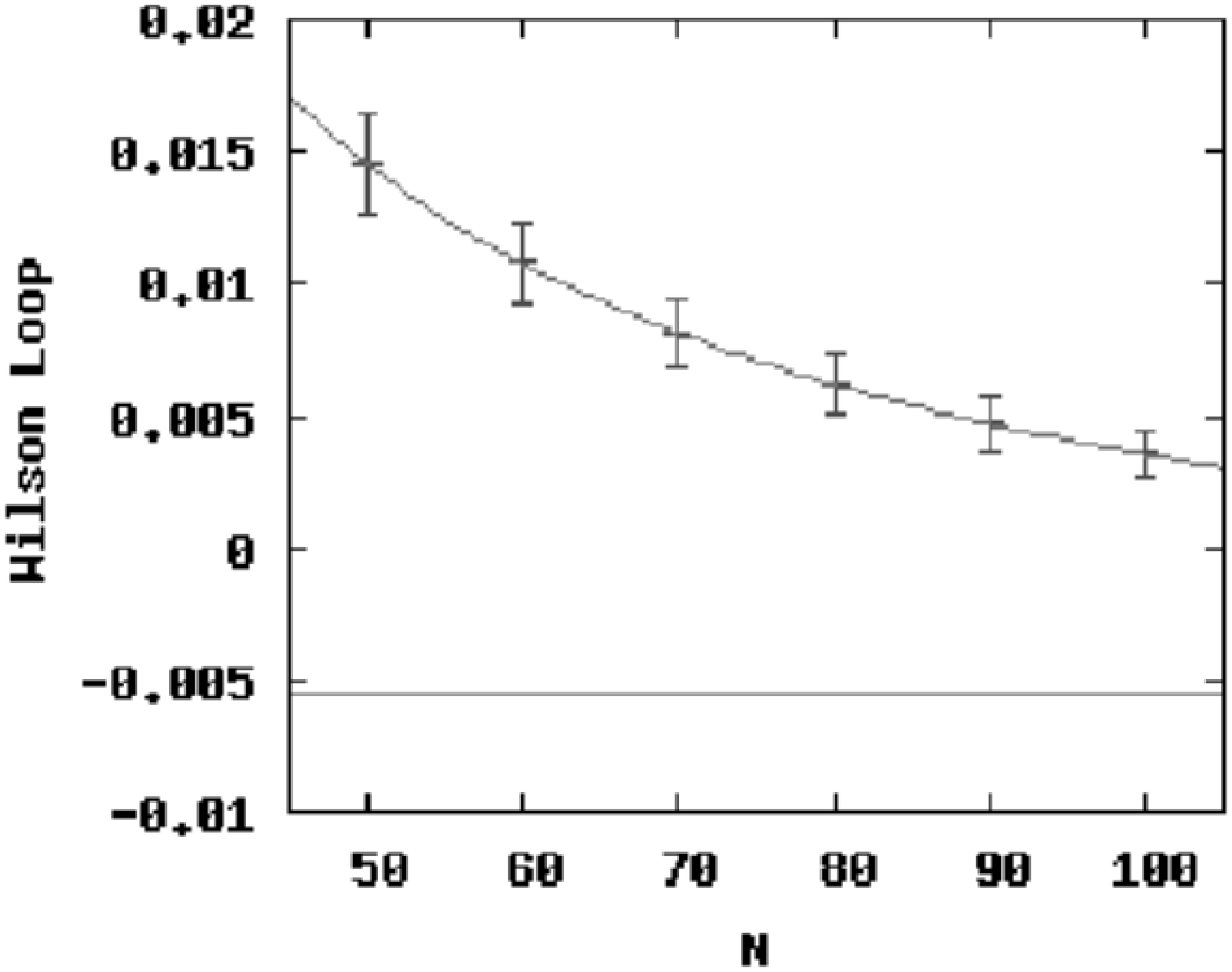}}
        \caption{
          $W[5,5]$ for $\alpha=3.73,\ \beta=0.4$. 
          The horizontal line represents the estimated value 
          in the large-$N$ limit, obtained by fitting the form
          $W[5,5]=c_1+c_2 N^{c_3}$, where $c_1,c_2$ and $c_3$ 
          are constants.
          We cannot find good convergence at this level. 
        }\label{Fig:N dependence of W[5,5]}
 }
\end{figure}
\subsubsection{$\alpha=0$: Original Weingarten model
}\label{subsec:original Weingarten}
\begin{figure}[htbp]
 \parbox{\halftext}{
   \scalebox{0.35}{\includegraphics{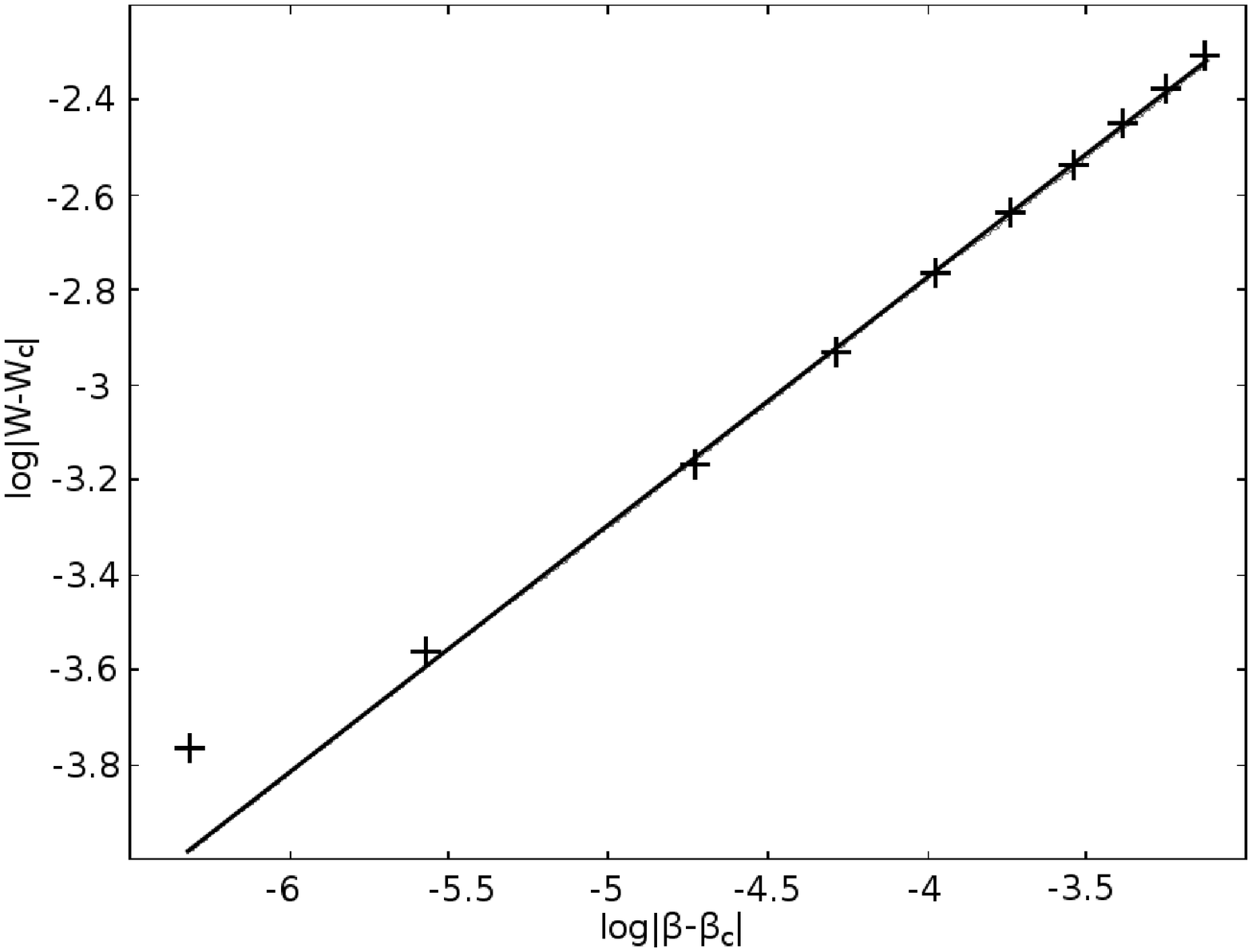}}
 }
 \hspace{5mm}
 \parbox{\halftext}{
   \scalebox{0.35}{\includegraphics{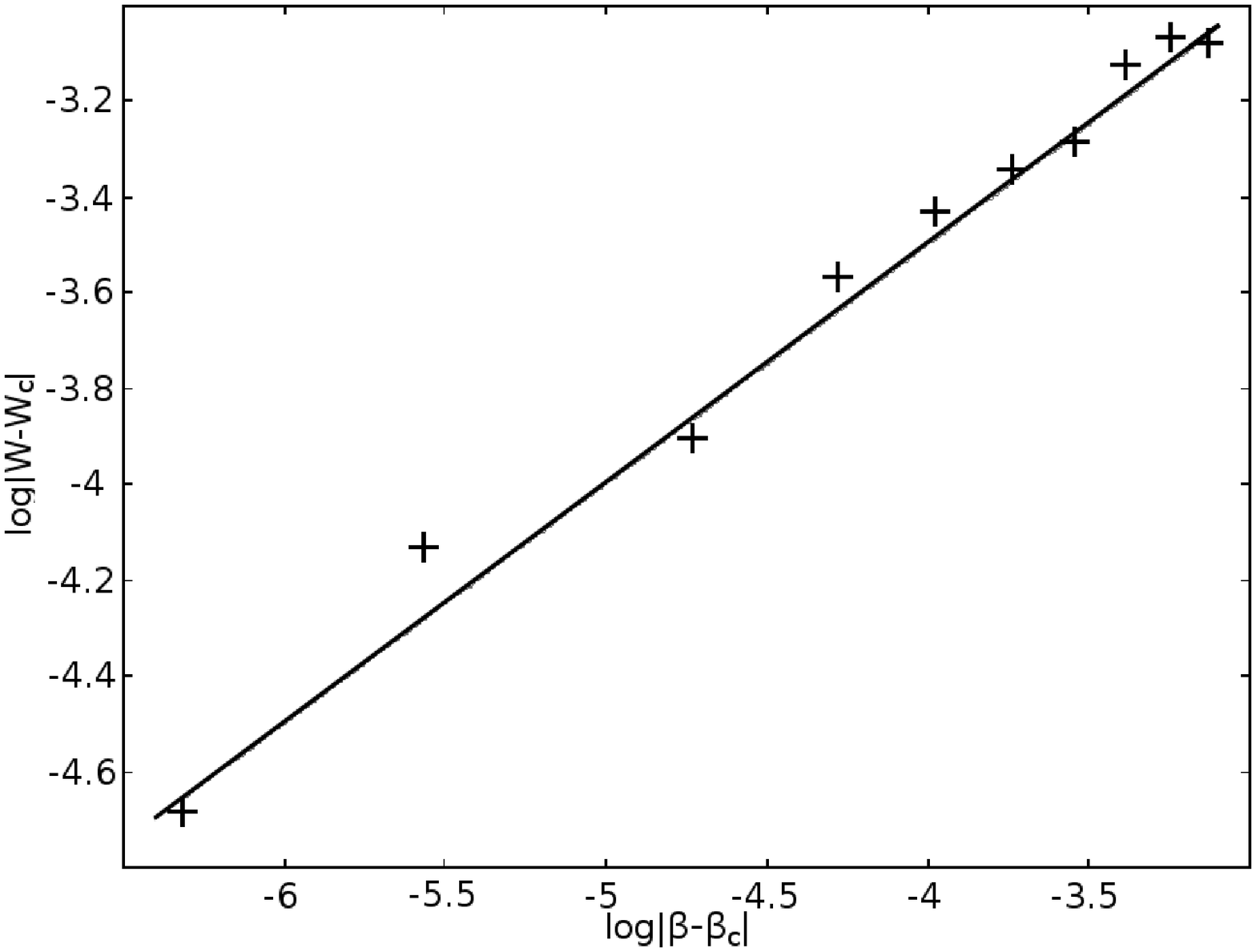}}
 }
 \caption{
      $\log|W_c[m,n]-W[m,n]|$ as a function of
      $\log|\beta-\beta_c|$ 
      for $\alpha=0$ (original Weingarten model), with $N=100$. 
      Here, the perimeter effect is taken into account. 
      The slope of the line is $0.5$. 
      [Left] $(m,n)=(1,1)$.  
      [Right] $(m,n)=(2,1)$.    
    }\label{fig:original Weingarten}
\end{figure}

For $\alpha=0$, it is known that  
\cite{DFJ84,KO}
\begin{eqnarray}
  b=-1.5.
  \qquad(\mbox{theoretical})
\end{eqnarray}
Then, taking the perimeter effect into account, 
we find that the numerical data are
consistent with this value 
(Fig.~\ref{fig:original Weingarten}). 
From the value of $W[1,1]$, we obtained 
\begin{eqnarray}
  b=-1.50\pm 0.02. 
  \qquad
  (\mbox{numerical})
\end{eqnarray}

\subsubsection{$0<\alpha\lesssim 1.2$
}\label{subsec:small alpha}
\begin{figure}[t]
 \parbox{\halftext}{
    \scalebox{0.3}{
    \includegraphics{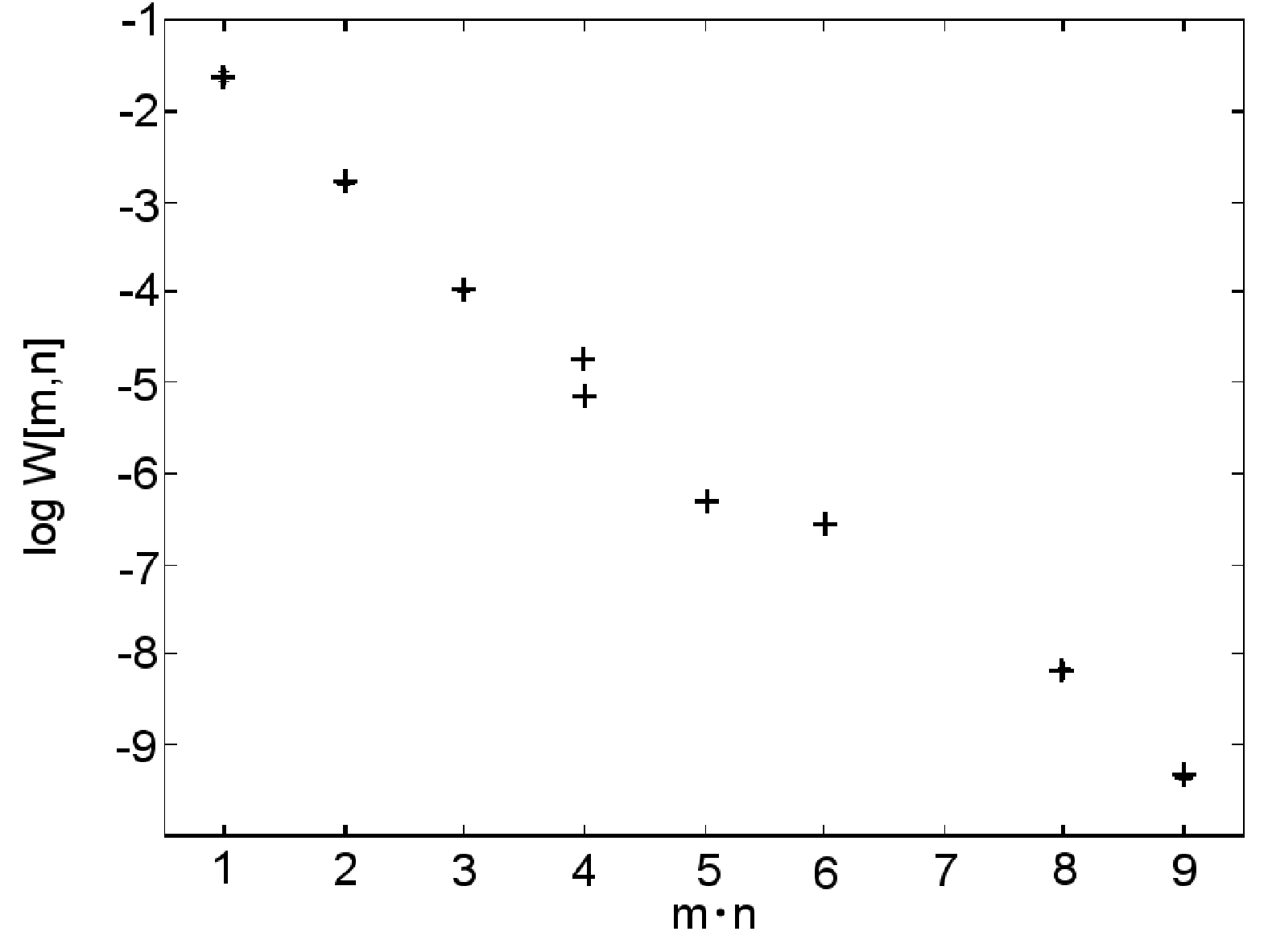}}
 }
 \hspace{4.5mm}
 \parbox{\halftext}{
    \scalebox{0.3}{
    \includegraphics{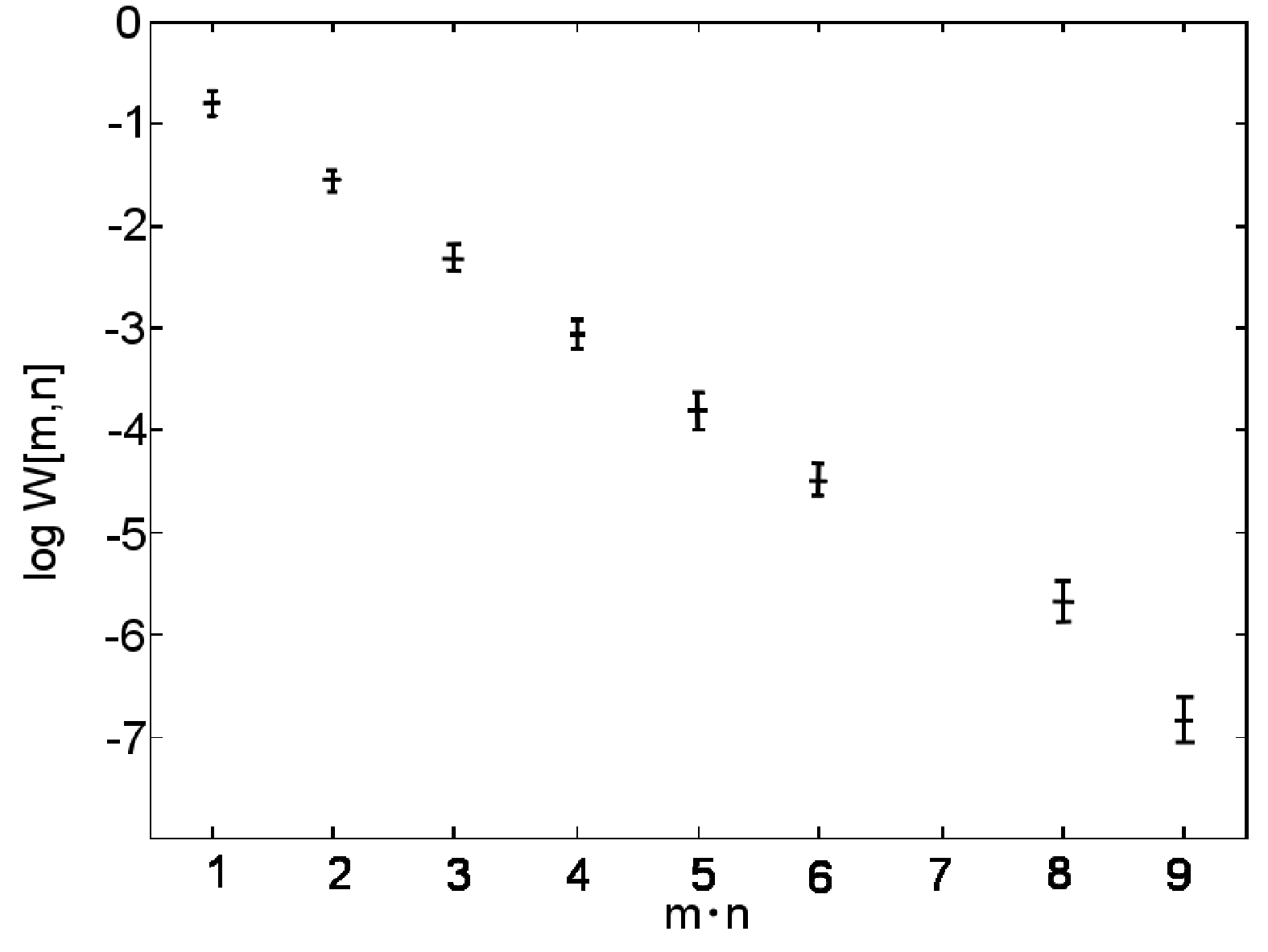}}
 }
    \caption{
      $\log W[m,n]$ as a function of the $A=mn$
      for $\alpha=1, N=100$. 
      [Left]
      Without the perimeter effect taken into consideration.
      [Right]
      With the perimeter effect taken into consideration. 
      The slope corresponds to $-T$. 
      The error bars become larger due to ambiguity in $p$. 
  }\label{fig:area law}
\end{figure}

\begin{figure}[t]
 \parbox{\halftext}{
        \scalebox{0.35}{\includegraphics{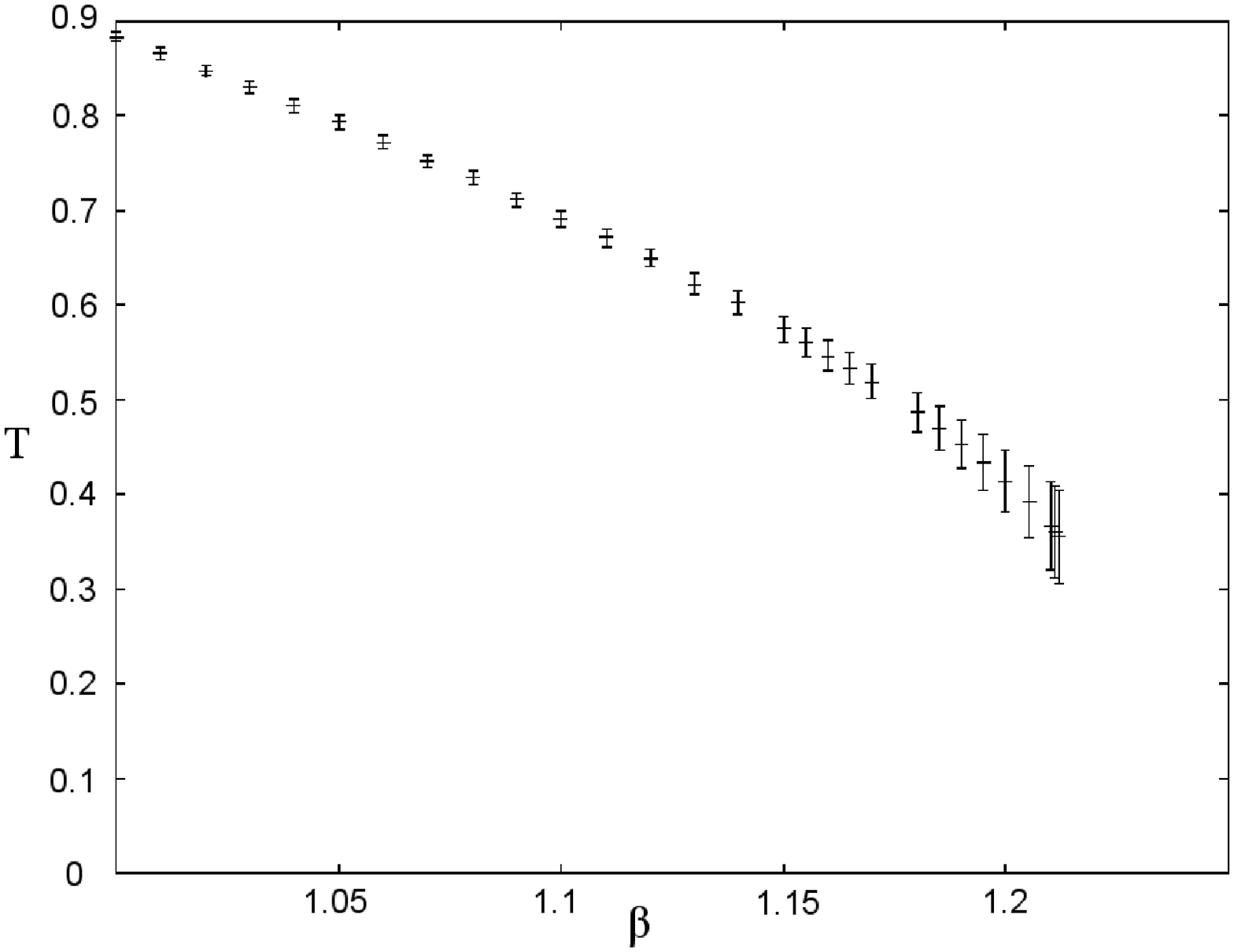}}
        \caption{
          The string tension $T$ for $\alpha=1.2, N=100$. 
        }\label{fig:tension_a=1.2}
 }
 \hspace{5mm}
 \parbox{\halftext}{
        \scalebox{0.35}{\includegraphics{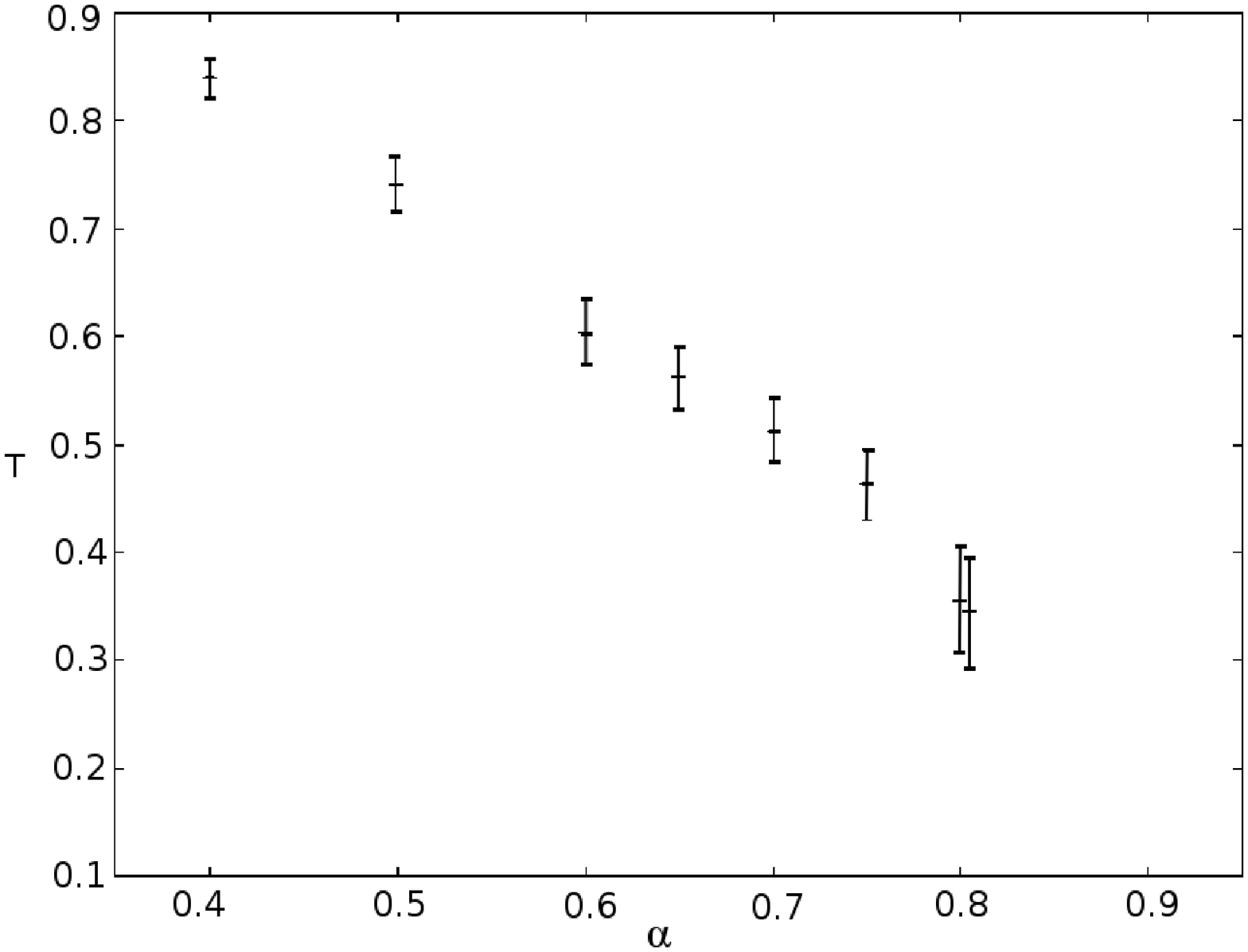}}
        \caption{
          The string tension $T$ for $\beta=\beta_c(\alpha), N=100$ 
          as a function of $\alpha$. 
        }\label{fig:tension} 
 }
\end{figure}

\begin{figure}[t]
 \parbox{\halftext}{
    \scalebox{0.35}{
    \includegraphics{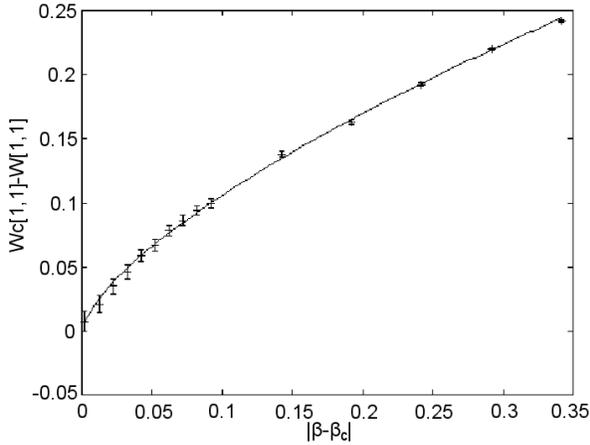}}
 }
 \hspace{5mm}
 \parbox{\halftext}{
    \scalebox{0.35}{
    \includegraphics{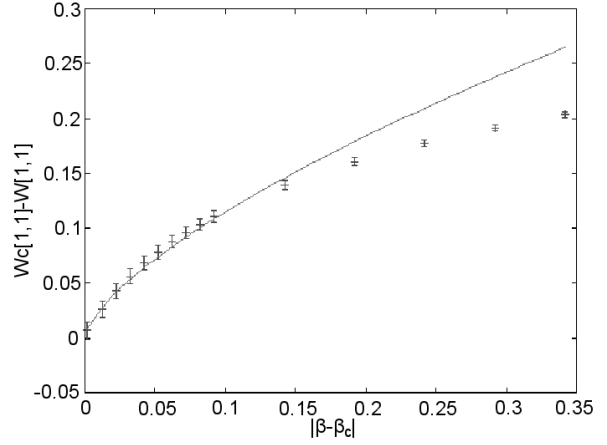}}
 }
    \caption{
      $W_c[1,1]-W[1,1]$ is as a function of $\beta_c-\beta$
      for $\alpha=1, N=100$. 
      The curves represent the fitted form 
      $W=W_c-\mbox{const}\cdot|\beta-\beta_c|^{-b-1}$. 
      [Left]
      With the perimeter effect taken into consideration.
      [Right]
      Without the perimeter effect taken into consideration.
      }\label{fig:susceptibility_1}
\end{figure}

\begin{figure}[t]
 \parbox{\halftext}{
        \scalebox{0.35}{ \includegraphics{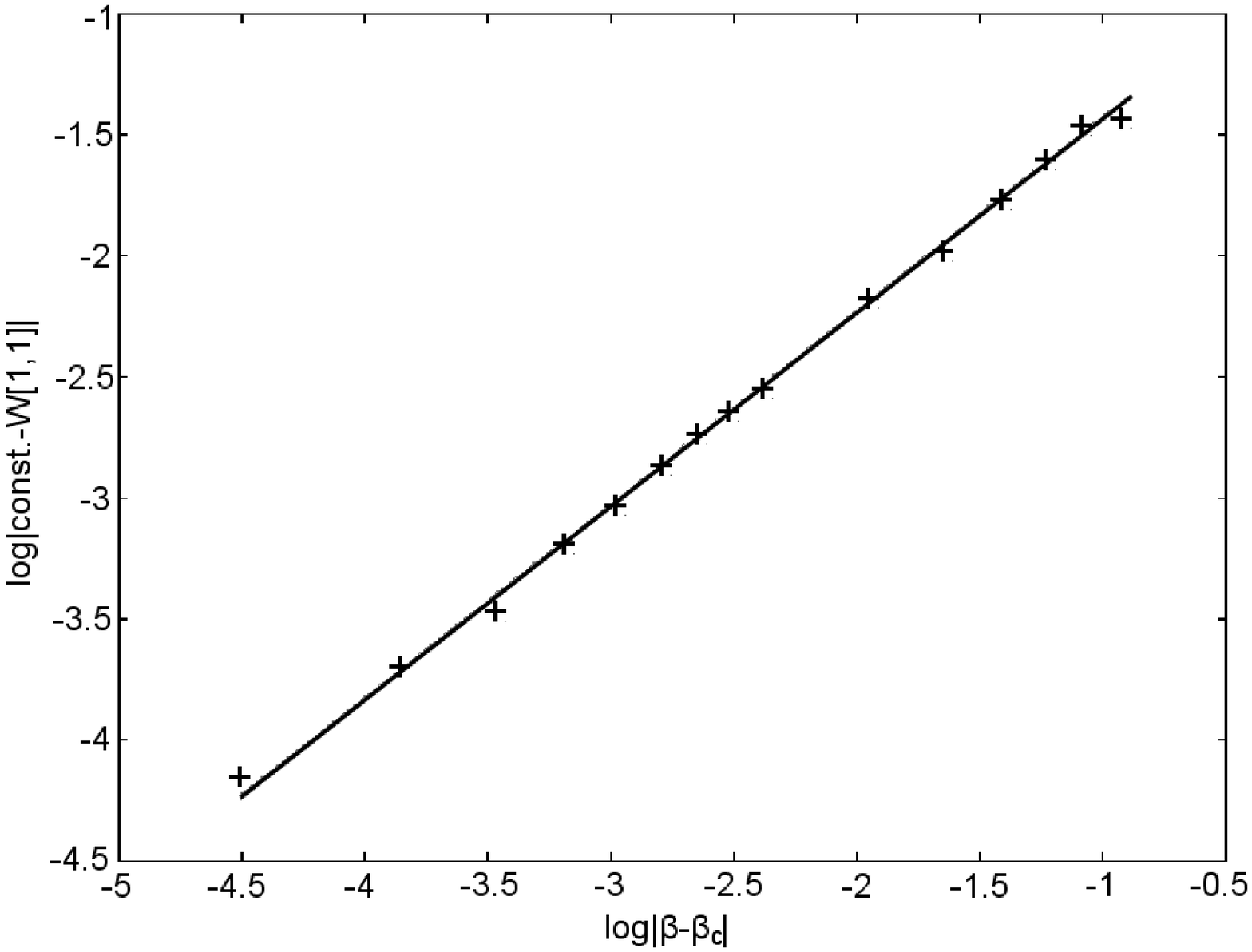}}
      \caption{
        $\log|W_c[1,1]-W[1,1]|$ as a function of $\log|\beta-\beta_c|$
        for $\alpha=1, N=100$.
        Here, the perimeter effect is taken into consideration.
        The line represents the fitted form, 
        whose slope is $-b-1$. 
      }\label{fig:susceptibility_2}
 }
 \hspace{4.5mm}
 \parbox{\halftext}{
        \scalebox{0.35}{\includegraphics{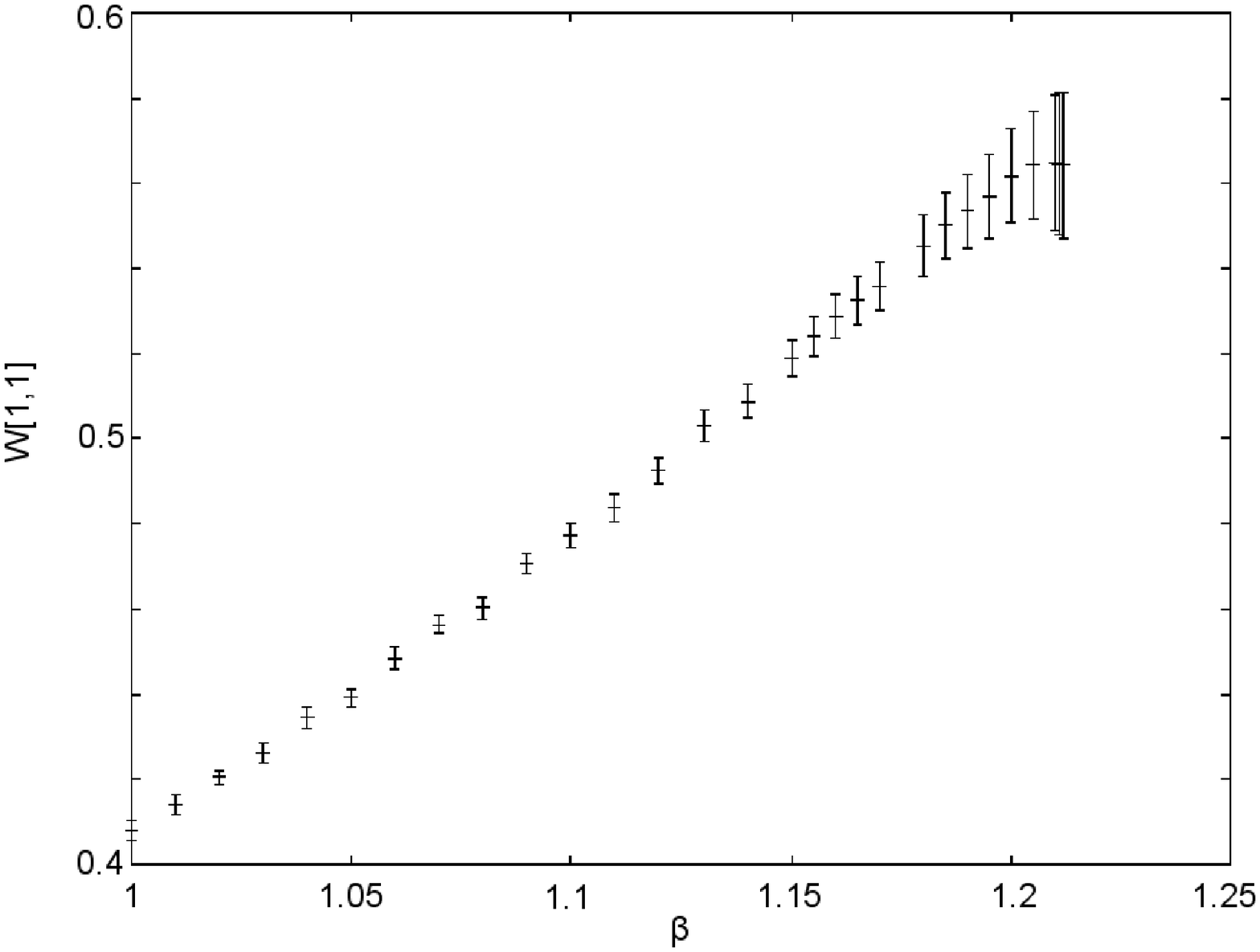}}
        \caption{
         $W[1,1]$ as a function of $\beta$
        for $\alpha=1.2, N=100$.
        Here, the perimeter effect is taken into account.     
        }\label{fig:W[1,1]}
 }
\end{figure}

\begin{figure}[t]
  \begin{center}
    \scalebox{0.35}{\includegraphics{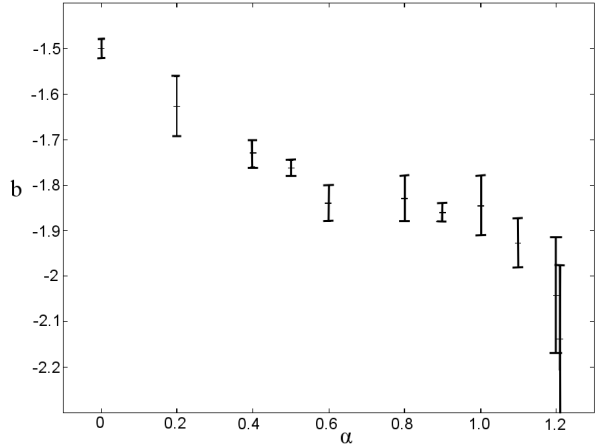}}
    \caption{
      String susceptibility $b$ as a function of $\alpha$.
      The result, $b=-1.50\pm 0.02$ at $\alpha=0$,
      is consistent with the result found
      in Refs.~\citen{DFJ84}, and \citen{KO}. 
    }\label{fig:susceptibility_3}
  \end{center}
\end{figure}

\noindent
\textbf{String tension}
\\
Taking the perimeter effect into account, we can determine 
the string tension $T$ (see Fig.~\ref{fig:area law}). 
For fixed $\alpha$, the string tension $T$ decreases 
as $\beta\to\beta_c(\alpha)$ (Fig.~\ref{fig:tension_a=1.2}). 
Although the value $T\left(\alpha,\beta_c(\alpha)\right)$   
decreases as $\alpha$ becomes large, it remains positive 
(see Fig.~\ref{fig:tension}).  
{}\\
\noindent
\textbf{String susceptibility}
\\
We next determine the ``string susceptibility'' $b(\alpha)$ 
using the numerical data and the expression 
(\ref{eq:expectation_value_of_Wilson_loop_with_perimeter_correction}). 
As can be seen from Fig.~\ref{fig:susceptibility_1}, the ansatz 
(\ref{eq:expectation_value_of_Wilson_loop_with_perimeter_correction}) 
agrees better with the data than the ansatz 
(\ref{eq:expectation_value_of_Wilson_loop_general_kappa}). 
Figure~\ref{fig:susceptibility_2} displays a plot of 
$\log|W_c[1,1]-W[1,1]|$ as a function of
$\log|\beta-\beta_c|$ for $\alpha=1$. 
We can read off the value
$-b-1$ from the slope of the fitting line.   
The value of $b(\alpha)$ is plotted in 
Fig.~\ref{fig:susceptibility_3}. For $\alpha\sim 1.2$, 
$b(\alpha)$ seems to approach $-2$ (Fig.~\ref{fig:susceptibility_3}).  
The uncertainties here are rather large,
and they come mainly from an ambiguity 
in the perimeter effect $p$. 
One of the reasons that they are so large 
is that the deviation of 
the value of the Wilson loop (at $N=100$) 
from that in the large-$N$ limit is not uniform but 
depends on the size of the loop and the parameters $\alpha$ and
$\beta$. 
If we can use larger matrix size $N$, 
we would be able to make the uncertainties smaller.  
\subsubsection{$1.2\lesssim\alpha<\infty$
}\label{subsec:large alpha}
In this parametric region, a phase transition takes place 
at $\beta=\beta_\mathrm{breakdown}<\beta_c$. 
As remarked in \textsection~\ref{sec:Phase diagram}, 
the $U(1)^2$ symmetry seems to be broken 
for $\beta>\beta_\mathrm{breakdown}$. 
Therefore, it is plausible that 
the reduced model deviates from that on a lattice, 
and thus the lattice-string interpretation is not
possible.\footnote{
  Similar phenomena in the unitary gauge theory are
  studied in Ref.~\citen{GNN}.  
}
\section{Conclusions and discussions}\label{sec:Discussion}
In this paper, we have
reported the results of a
a numerical study of the two-dimensional generalized 
Weingarten model (\ref{action}).  
If we assume the relations 
(\ref{ansatz:suppression of number of random surface}) and 
(\ref{eq:expectation_value_of_Wilson_loop_with_perimeter_correction}), 
then the numerical data show that the string susceptibility 
approaches $-2$ as we increase the value of the parameter 
$\alpha$ to $\sim 1.2$. 
This result suggests that branched-polymer configurations 
are suppressed in this parametric region. 
We also found that the string tension decreases. 
However, we encountered a phase transition before 
the string tension becomes zero. Therefore, we cannot 
take the continuum limit in the reduced model studied in this paper. 

There are several pontential directions for future studies. 
First, it is necessary to understand the present model in terms of 
the random surface and explain why 
the assumption (\ref{ansatz:suppression of number of random surface}) 
seems to be consistent with the numerical data. 
Doing so, we would be able to clarify 
whether or not our numerical result actually indicates that 
the branched polymer configurations are suppressed. 
In addition, such an understanding would be helpful 
for finding better models. 
Second, it would be interesting to study the model 
defined on a lattice \cite{IMS}. 
In the case of the large-$N$ reduced $d$-dimensional 
$U(N)$ gauge theory ($d\ge 3$), the breakdown of $U(1)^d$ 
is an artifact of the reduced model;  
if the model is defined on 
a lattice of size $L^d$, then the $U(1)^d$ symmetry 
is not broken to the weaker coupling 
as $L$ increases \cite{NN}.  
If similar phenomena exist in the present case, 
then using the model with $L>1$, we could study the larger parametric 
region, and we may be able to
find a point at which we can take the continuum limit. 
Third, we can also consider higher-dimensional models. 
In this case, the lattice-string interpretation may be valid
in the parametric region in which some of the $U(1)$s 
remain unbroken.
This point is worth studying. We hope to report analysis of 
these models in future publications.

\section*{Acknowledgements}
THe numerical computations used in this work were carried out at
the Yukawa Institute Computer Facility. 
The authors thank Hikaru Kawai and Takashi Kanai 
for stimulating discussions and comments. 
A part of our simulation code is that used in 
collaboration with them \cite{HKKK}. 
M.~H. also thanks Tatsuo Azeyanagi, Tomoyoshi Hirata and 
Yoshinori Matsuo for useful discussions.  
M.~H. would like to 
thank the Japan Society for the Promotion of Science for financial
support. 
He was also supported in part by the JSPS and the French Ministry of 
Foreign Affairs under the Japan-France Integrated Action 
Program (SAKURA). 
He would like to thank CEA/Saclay and Ivan Kostov for hospitality.
F.~K. was supported in part by a Grant-in-Aid for
the 21st Century COE ``Center for Diversity and Universality in
Physics''. 

\appendix
\section{Comparison with the Two-Dimensional 
Generalized Weingarten Model without a Twist}\label{appendix:untwieted}
\subsection{Two-dimensional generalized Weingarten model without a twist}
\begin{figure}[htbp]
 \parbox{\halftext}{
        \scalebox{0.38}{\input{phase_small_alpha_nontwist.pstex_t}}
        \caption{
          Phase diagram of the two-dimensional generalized Weingarten
          model {\it without a twist} 
          for $\alpha\lesssim 1$. 
        }\label{phase diagram: small alpha without twist}
 }
 \hspace{-1.5mm}
 \parbox{\halftext}{
        \scalebox{0.38}{\input{phase_large_alpha.pstex_t}}
        \caption{
          Phase diagram of the two-dimensional generalized Weingarten
          model {\it without a twist} 
          for $\alpha\gtrsim 1$. (This figure is based on Fig.~10 of Ref.~\citen{HKKK}.)  
        }\label{phase diagram: large alpha without twist}
 }
\end{figure}
In this section, we consider 
the original, untwisted generalized Weingarten model 
in two dimensions.   

For large, fixed $\alpha$ there are two curves of first-order phase
transitions.
We call them $\beta_1$ and $\beta_2$ in ascending order. 
They correspond to the breakdown of the $U(1)^2$ symmetry. 
If we increase $\beta$ with $\alpha$ fixed, 
$U(1)^2$ is broken to $U(1)$ at $\beta_1$,
and then it is broken completely at $\beta_2$. 
The values of $\beta_1$ and $\beta_2$ 
seem to diverge as $\alpha^{-1}\to 0$. 
This is consistent with the analytic result, 
in which $U(1)^2$ is not broken.
At small $\alpha$, $\beta_1$ and $\beta_2$ seem to become equal. 
This transition persists to $\alpha\sim 1.2$ where it merges with  
the boundary of the metastable parametric region. 
Near the boundary of 
the well-defined region $\beta<\alpha$, 
the $U(1)^2$ symmetry is restored. 
In Fig.~\ref{phase diagram: small alpha without twist}, 
this line of restoration of $U(1)^2$ is drawn by a dotted line 
near $\alpha\sim 1.2$,   
because the restoration cannot be seen clearly in this region. 

In the parametric region where the $U(1)^2$ symmetry is not broken, 
our reduced model is equivalent to the model defined on a lattice 
\cite{IMS} through the large-$N$ reduction. 
If this symmetry is broken, then 
the reduced model deviates from that on a lattice, 
and the lattice-string interpretation is not valid
\footnote{
  For $d\ge 3$, if $U(1)^{d^\prime}$ remains unbroken, 
  then this model is equivalent to a model defined 
  on a $d^\prime$-dimensional lattice coupled to 
  $d-d^\prime$ adjoint scalars. 
  For this reason, such a parametric region is of interest in this case.  
  Note that the $U(1)$s do not necessarily break one-by-one 
  in the case of the twisted model.  
}. 
\subsection{Comparison of untwisted and twisted models}\label{appendix:maximal twist}
\begin{figure}[t]
 \parbox{\halftext}{
        \scalebox{0.28}{\includegraphics{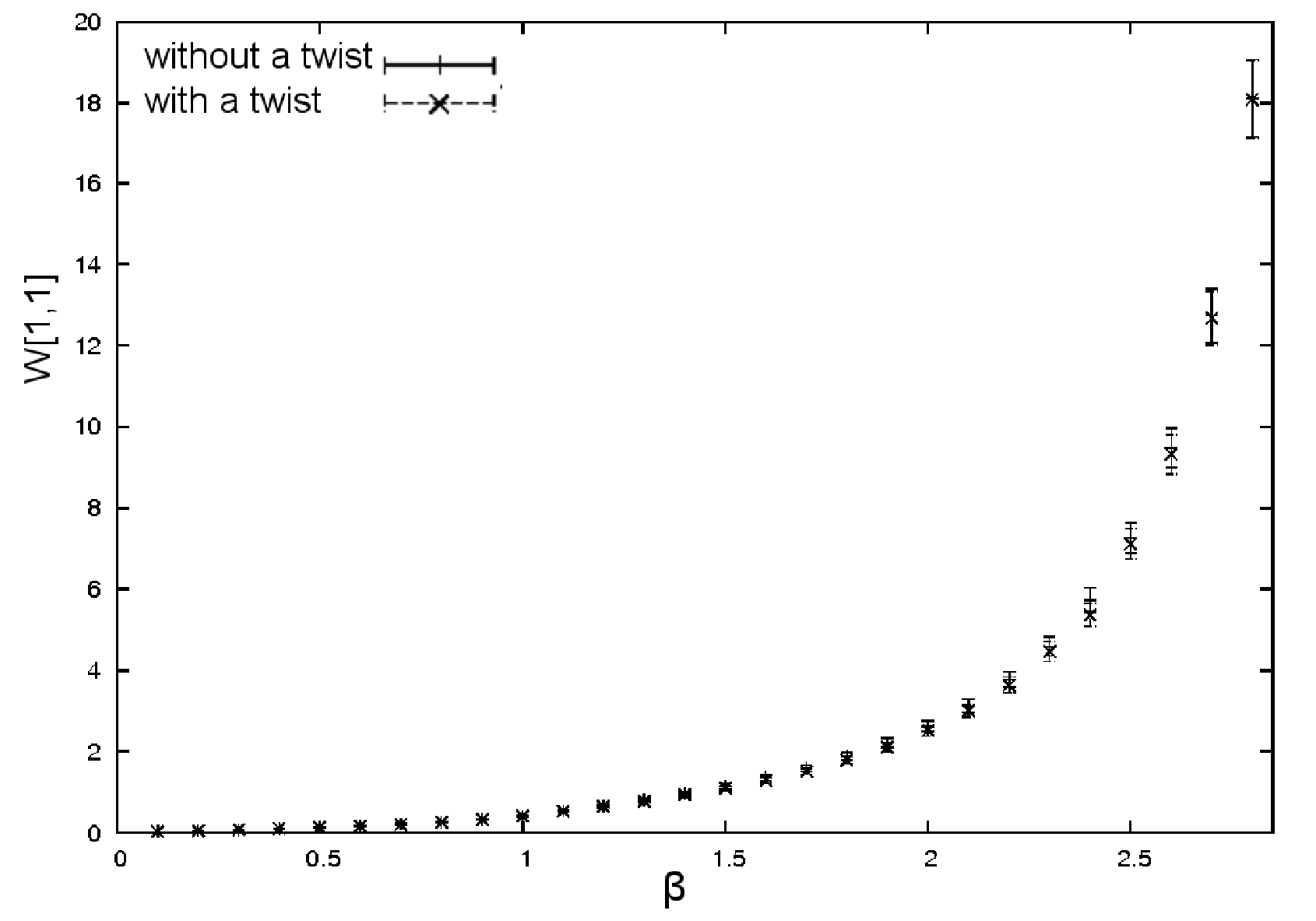}}
      \caption{
        Plot of the expectation value of $1\times 1$ 
        Wilson loop for $\alpha^{-1}=0.3, N=50$.
      }\label{Fig:Wilson_AI=3}
 }
 \hspace{1.5mm}
 \parbox{\halftext}{
        \scalebox{0.28}{\includegraphics{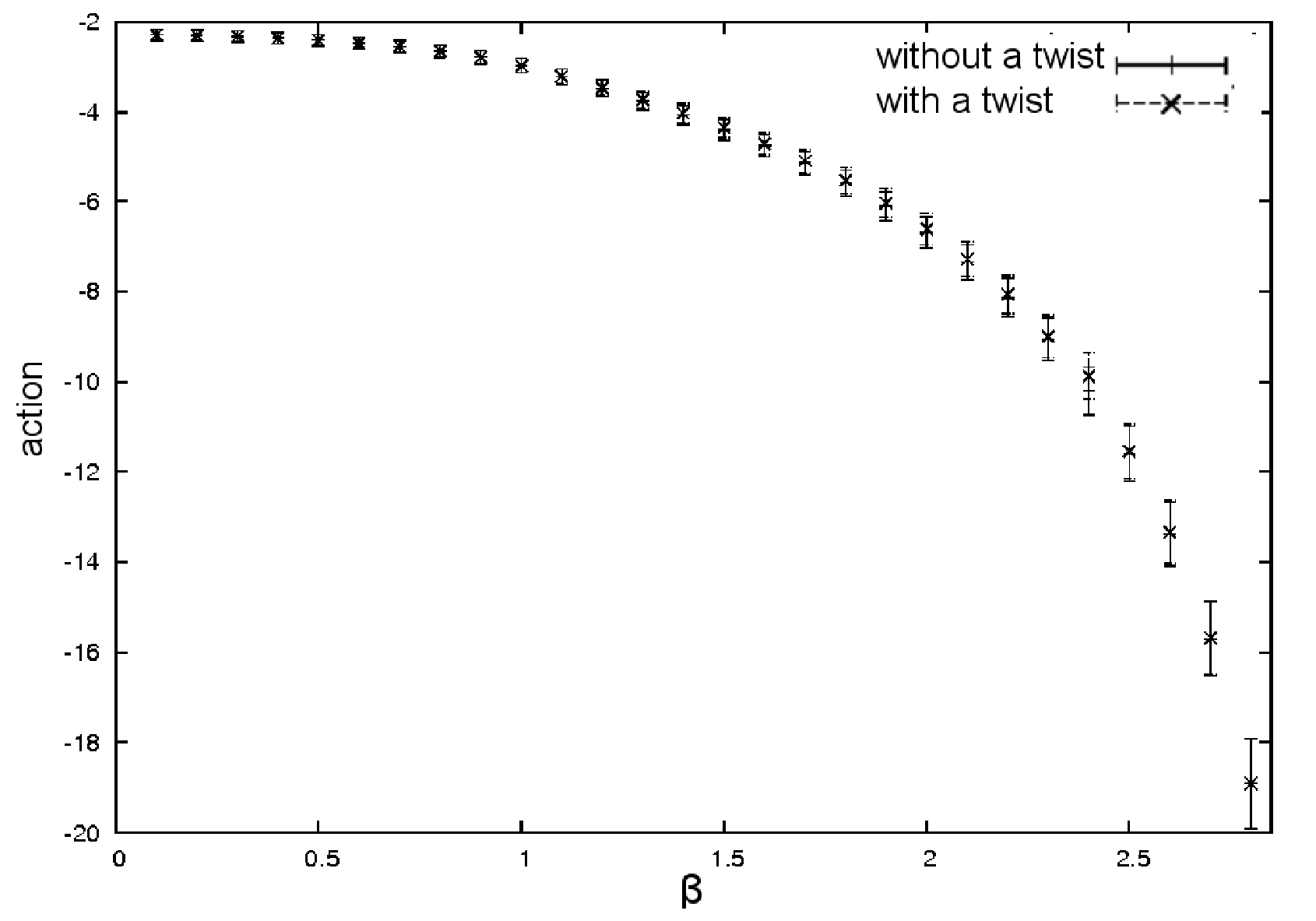}}
        \caption{ 
          Plot of the expectation value of the action 
          for $\alpha^{-1}=0.3, N=50$. 
        }\label{Fig:Action_AI=3}
 }
\end{figure}
In this subsection, in order to avoid confusion, 
we indicate the physical quantities in the twisted model 
with the subscript $T$. For example, Wilson loops are denoted as 
\begin{eqnarray}
  & &
  W[m,n](\alpha,\beta),
  \quad
  \quad
  (\mbox{without twist})
  \\
  & &
  W_T[m,n](\alpha,\beta)
  =
  (-)^{mn}W[m,n](\alpha,-\beta).
  \quad
  \quad
  (\mbox{with maximal twist}) 
\end{eqnarray}
Here we have included the phase $(-)^{mn}$ so that 
the two prescriptions give the same value
in the strong coupling region: 
\begin{eqnarray}
  W[m,n](\alpha,\beta)
  =
  W_T[m,n](\alpha,\beta).   
  \qquad
  (\mbox{strong coupling})
\end{eqnarray}
In Fig.~\ref{Fig:Wilson_AI=3}, we plot the expectation values 
of $1\times 1$ Wilson loops for $\alpha^{-1}=0.3$. 
We see that $W[1,1]$ and $W_T[1,1]$ indeed take the same value. 
We can also see that the expectation values of the action 
are nearly equal (Fig.~\ref{Fig:Action_AI=3}). 

\begin{figure}[t]
 \parbox{\halftext}{
        \scalebox{0.28}{\includegraphics{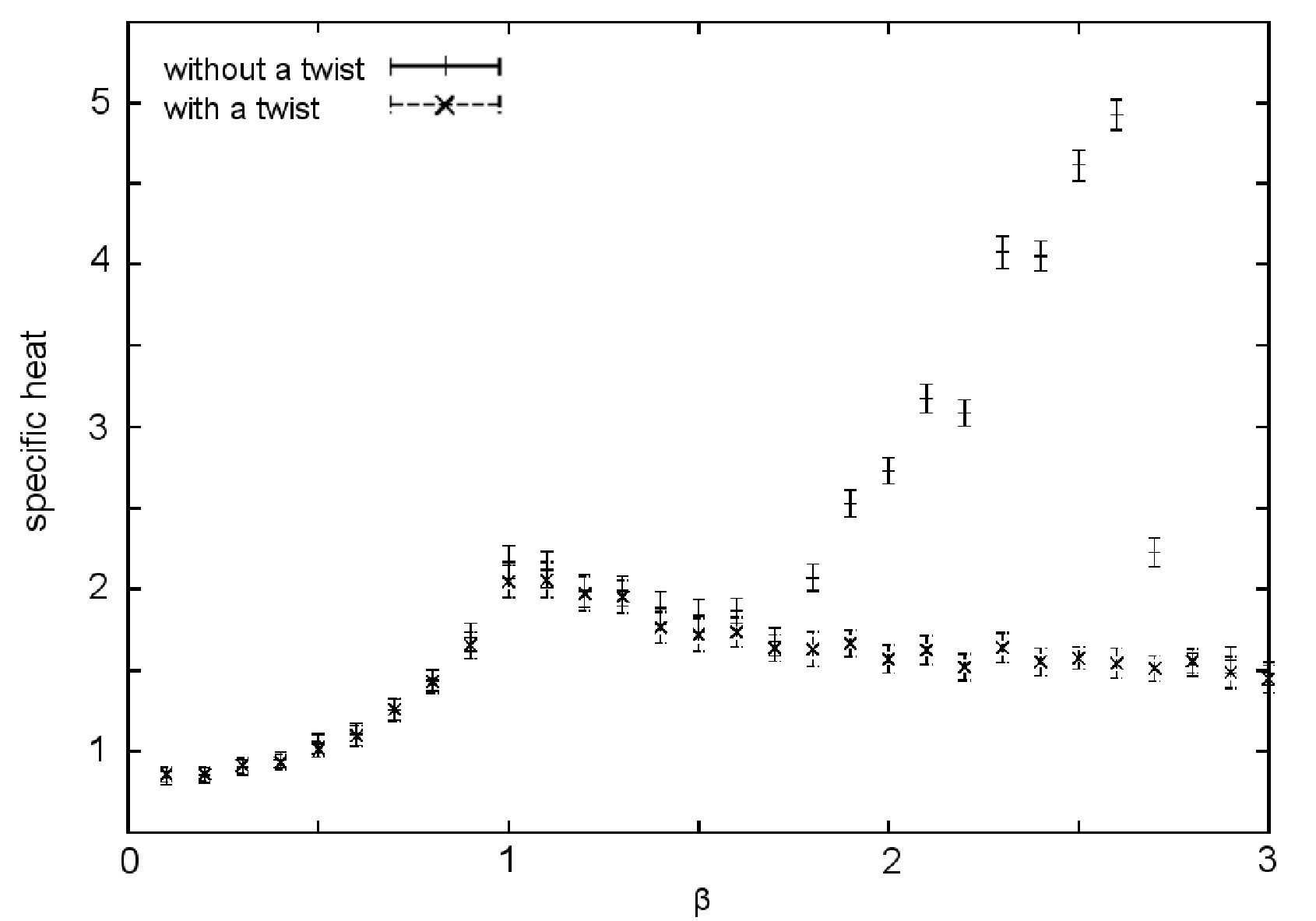}}
        \caption{ 
          Plot of the specific heat for $\alpha^{-1}=0.3, N=50$. 
        }\label{Fig:SpecificHeat_AI=3}
 }
 \hspace{4.5mm}
 \parbox{\halftext}{
        \scalebox{0.28}{\includegraphics{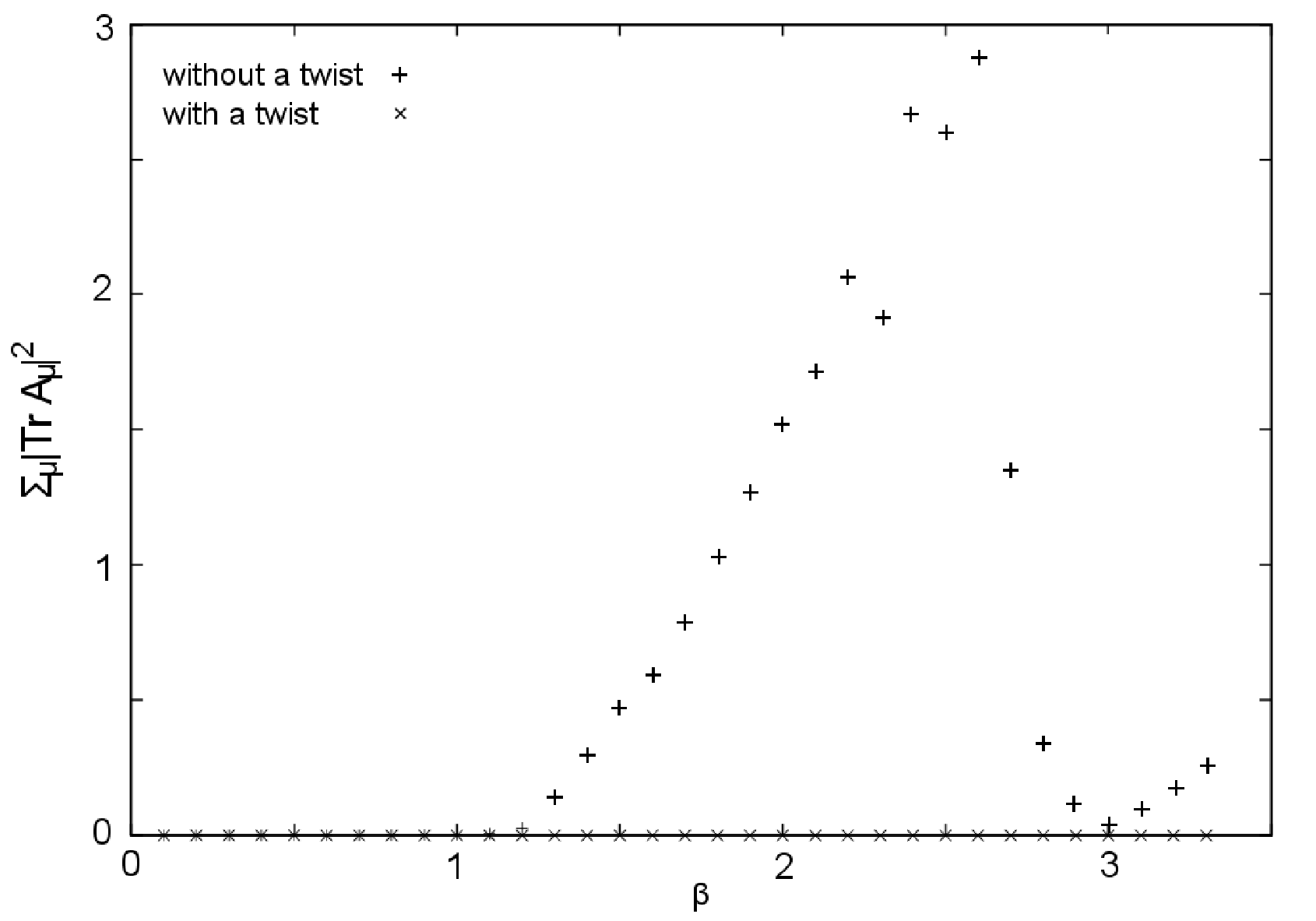}}
      \caption{
        Plot of the expectation value of 
        $\sum_{\mu}|\Tr A_\mu|^2$ for $\alpha^{-1}=0.3, N=50$. 
      }\label{Fig:Trace_AI=3}
 }
\end{figure}

It is interesting that the
two prescriptions seem to give the same expectation values 
for the Wilson loops not only in the strong coupling region 
but also in the weak coupling region, which is separated from 
the strong coupling region by a phase transition;  
indeed, as can be seen from Fig.~\ref{Fig:SpecificHeat_AI=3}, 
phase transitions take place naer $\beta_1(\alpha^{-1}=0.3)= 1$ 
both in the twisted and untwisted models. 
In the case of the untwisted model, this transition corresponds to 
the breakdown of $U(1)^2$ (see Fig.~\ref{Fig:Trace_AI=3}). 
In the case of the twisted model, because the expectation value 
of $\sum_{\mu}|\Tr A_\mu|$ remains nearly
equal to zero, this transition 
does not correspond to the complete breakdown of one of the $U(1)$s;  
it may represent a breakdown to ${\mathbb Z}_n$ for some integer 
$n$.\footnote{A similar conjecture is made in Ref.~\citen{GNN}
in the case of 
the four-dimensional twisted Eguchi-Kawai model.} 

\end{document}

%% file: phase_small_alpha.pstex_t
\begin{picture}(0,0)%
\includegraphics{phase_small_alpha.pstex}%
\end{picture}%
\setlength{\unitlength}{3947sp}%
\begingroup\makeatletter\ifx\SetFigFont\undefined%
\gdef\SetFigFont#1#2#3#4#5{%
  \reset@font\fontsize{#1}{#2pt}%
  \fontfamily{#3}\fontseries{#4}\fontshape{#5}%
  \selectfont}%
\fi\endgroup%
\begin{picture}(7413,7362)(1,-7486)
\put(6751,-511){\makebox(0,0)[lb]{\smash{{\SetFigFont{29}{34.8}{\rmdefault}{\mddefault}{\updefault}{\color[rgb]{0,0,0}$\beta$}%
}}}}
\put(5326,-2836){\makebox(0,0)[lb]{\smash{{\SetFigFont{29}{34.8}{\rmdefault}{\mddefault}{\updefault}{\color[rgb]{0,0,0}$\beta=\beta_c$}%
}}}}
\put(976,-7111){\makebox(0,0)[lb]{\smash{{\SetFigFont{29}{34.8}{\familydefault}{\mddefault}{\updefault}{\color[rgb]{0,0,0}$\alpha$}%
}}}}
\put(1576,-3211){\makebox(0,0)[lb]{\smash{{\SetFigFont{34}{40.8}{\rmdefault}{\mddefault}{\updefault}{\color[rgb]{0,0,0}stable}%
}}}}
\put(2401,-1186){\makebox(0,0)[lb]{\smash{{\SetFigFont{34}{40.8}{\rmdefault}{\mddefault}{\updefault}{\color[rgb]{0,0,0}metastable}%
}}}}
\put(7051,-6136){\makebox(0,0)[lb]{\smash{{\SetFigFont{29}{34.8}{\rmdefault}{\mddefault}{\updefault}{\color[rgb]{0,0,0}$\alpha=\beta$}%
}}}}
\put(  1,-3886){\makebox(0,0)[lb]{\smash{{\SetFigFont{29}{34.8}{\rmdefault}{\mddefault}{\updefault}{\color[rgb]{0,0,0}    $\sim 1.2$}%
}}}}
\put(4426,-6811){\makebox(0,0)[lb]{\smash{{\SetFigFont{29}{34.8}{\rmdefault}{\mddefault}{\updefault}{\color[rgb]{0,0,0}$\beta=\beta_{breakdown}$}%
}}}}
\put(5476,-7486){\makebox(0,0)[lb]{\smash{{\SetFigFont{29}{34.8}{\rmdefault}{\mddefault}{\updefault}{\color[rgb]{0,0,0}first order?}%
}}}}
\end{picture}%

%% file: phase_small_alpha_nontwist.pstex_t
\begin{picture}(0,0)%
\includegraphics{phase_small_alpha_nontwist.pstex}%
\end{picture}%
\setlength{\unitlength}{3947sp}%
\begingroup\makeatletter\ifx\SetFigFont\undefined%
\gdef\SetFigFont#1#2#3#4#5{%
  \reset@font\fontsize{#1}{#2pt}%
  \fontfamily{#3}\fontseries{#4}\fontshape{#5}%
  \selectfont}%
\fi\endgroup%
\begin{picture}(7212,7422)(1,-7546)
\put(6751,-511){\makebox(0,0)[lb]{\smash{{\SetFigFont{29}{34.8}{\rmdefault}{\mddefault}{\updefault}{\color[rgb]{0,0,0}$\beta$}%
}}}}
\put(5326,-2836){\makebox(0,0)[lb]{\smash{{\SetFigFont{29}{34.8}{\rmdefault}{\mddefault}{\updefault}{\color[rgb]{0,0,0}$\beta=\beta_c$}%
}}}}
\put(976,-7111){\makebox(0,0)[lb]{\smash{{\SetFigFont{29}{34.8}{\familydefault}{\mddefault}{\updefault}{\color[rgb]{0,0,0}$\alpha$}%
}}}}
\put(1576,-3211){\makebox(0,0)[lb]{\smash{{\SetFigFont{34}{40.8}{\rmdefault}{\mddefault}{\updefault}{\color[rgb]{0,0,0}stable}%
}}}}
\put(2401,-1186){\makebox(0,0)[lb]{\smash{{\SetFigFont{34}{40.8}{\rmdefault}{\mddefault}{\updefault}{\color[rgb]{0,0,0}metastable}%
}}}}
\put(7051,-6136){\makebox(0,0)[lb]{\smash{{\SetFigFont{29}{34.8}{\rmdefault}{\mddefault}{\updefault}{\color[rgb]{0,0,0}$\alpha=\beta$}%
}}}}
\put(  1,-3886){\makebox(0,0)[lb]{\smash{{\SetFigFont{29}{34.8}{\rmdefault}{\mddefault}{\updefault}{\color[rgb]{0,0,0}    $\sim 1.2$}%
}}}}
\put(3901,-6511){\makebox(0,0)[lb]{\smash{{\SetFigFont{29}{34.8}{\rmdefault}{\mddefault}{\updefault}{\color[rgb]{0,0,0}first order?}%
}}}}
\put(3676,-6061){\makebox(0,0)[lb]{\smash{{\SetFigFont{29}{34.8}{\rmdefault}{\mddefault}{\updefault}{\color[rgb]{0,0,0}$\beta=\beta_1=\beta_2$}%
}}}}
\put(6226,-7411){\makebox(0,0)[lb]{\smash{{\SetFigFont{29}{34.8}{\rmdefault}{\mddefault}{\updefault}{\color[rgb]{0,0,0}$\{1\}\to U(1)^2$}%
}}}}
\end{picture}%

%% file: phase_large_alpha.pstex_t
\begin{picture}(0,0)%
\includegraphics{phase_large_alpha.pstex}%
\end{picture}%
\setlength{\unitlength}{3947sp}%
\begingroup\makeatletter\ifx\SetFigFont\undefined%
\gdef\SetFigFont#1#2#3#4#5{%
  \reset@font\fontsize{#1}{#2pt}%
  \fontfamily{#3}\fontseries{#4}\fontshape{#5}%
  \selectfont}%
\fi\endgroup%
\begin{picture}(7725,6852)(76,-6853)
\put(4351,-2836){\makebox(0,0)[lb]{\smash{{\SetFigFont{29}{34.8}{\rmdefault}{\mddefault}{\updefault}{\color[rgb]{0,0,0}unstable}%
}}}}
\put(4651,-3811){\makebox(0,0)[lb]{\smash{{\SetFigFont{29}{34.8}{\rmdefault}{\mddefault}{\updefault}{\color[rgb]{0,0,0}$\alpha=\beta$}%
}}}}
\put(7351,-6661){\makebox(0,0)[lb]{\smash{{\SetFigFont{29}{34.8}{\rmdefault}{\mddefault}{\updefault}{\color[rgb]{0,0,0}$\beta$}%
}}}}
\put(226,-361){\makebox(0,0)[lb]{\smash{{\SetFigFont{29}{34.8}{\rmdefault}{\mddefault}{\updefault}{\color[rgb]{0,0,0}$\alpha^{-1}$}%
}}}}
\put( 76,-2011){\makebox(0,0)[lb]{\smash{{\SetFigFont{29}{34.8}{\rmdefault}{\mddefault}{\updefault}{\color[rgb]{0,0,0}$\sim 0.6$}%
}}}}
\put(7801,-6136){\makebox(0,0)[lb]{\smash{{\SetFigFont{29}{34.8}{\rmdefault}{\mddefault}{\updefault}{\color[rgb]{0,0,0}$\beta_1: U(1)^2\to U(1)$}%
}}}}
\put(7801,-5686){\makebox(0,0)[lb]{\smash{{\SetFigFont{29}{34.8}{\rmdefault}{\mddefault}{\updefault}{\color[rgb]{0,0,0}$\beta_2: U(1)\to \{1\}$}%
}}}}
\put(7801,-5236){\makebox(0,0)[lb]{\smash{{\SetFigFont{29}{34.8}{\rmdefault}{\mddefault}{\updefault}{\color[rgb]{0,0,0}$\{1\}\to U(1)^2$}%
}}}}
\put(1576,-6736){\makebox(0,0)[lb]{\smash{{\SetFigFont{29}{34.8}{\rmdefault}{\mddefault}{\updefault}{\color[rgb]{0,0,0}$0.5$}%
}}}}
\put( 76,-5086){\makebox(0,0)[lb]{\smash{{\SetFigFont{29}{34.8}{\rmdefault}{\mddefault}{\updefault}{\color[rgb]{0,0,0}$0.268$}%
}}}}
\put(2626,-6211){\makebox(0,0)[lb]{\smash{{\SetFigFont{29}{34.8}{\rmdefault}{\mddefault}{\updefault}{\color[rgb]{0,0,0}third order transition}%
}}}}
\end{picture}%